\documentclass[a4paper,11pt]{article}
\usepackage{pos}

\title{Light scalars in the Weinberg 3HDM potential with spontaneous CP violation} 
\ShortTitle{Weinberg 3HDM potential}

\author*[a]{Robin Plantey}
\author[b]{O. M. Ogreid}
\author[c]{P. Osland}
\author[d]{M. N. Rebelo}
\author[a]{M. Aa. Solberg}

\affiliation[a]{Department of Structural Engineering, NTNU,\\7491 Trondheim, Norway}
\affiliation[b]{Western Norway University of Applied Sciences,\\Postboks 7030, N-5020 Bergen, Norway}
\affiliation[c]{Department of Physics and Technology, University of Bergen,\\Postboks 7803, N-5020 Bergen, Norway}
\affiliation[d]{Centro de Física Teórica de Partículas – CFTP and Dept de Física Instituto Superior Técnico – IST, Universidade de Lisboa,\\ Av.  Rovisco Pais, P-1049-001 Lisboa, Portugal}

\emailAdd{Robin.Plantey@ntnu.no}
\emailAdd{omo@hvl.no}
\emailAdd{Per.Osland@uib.no}
\emailAdd{rebelo@tecnico.ulisboa.pt}
\emailAdd{Marius.Solberg@ntnu.no}

\abstract{The $\mathbb Z_2\times \mathbb Z_2$~-symmetric 3HDM potential, sometimes called Weinberg's 3HDM potential, can accommodate both explicit and spontaneous CP violation as well as natural flavour conservation and is hence relevant for studies of CP violation coming from a realistic extended Higgs sector. The model has an interesting regime, controlled by a single parameter, where it acquires an approximate global $U(1) \times U(1)$ symmetry which causes CP violating effects to be small and two pseudo-Goldstone bosons to be present in the scalar spectrum. From parameter space scans, we find that, in a realistic implementation featuring an SM-like Higgs boson, this model  predominantely leads to the existence of light additional Higgs bosons with a mass lower than the SM-like Higgs boson.}
\FullConference{%
  7th Symposium on Prospects in the Physics of Discrete Symmetries (DISCRETE 2020-2021)\\
  29th November - 3rd December 2021\\
 Bergen, Norway}

\begin{document}
\maketitle

\def\Prms#1#2{\left(P_{#1#2}\right)_\text{r.m.s}}

\section{Introduction}

Weinberg realized that one could secure strangeness and charm conservation within a gauge theory, and yet have  explicit CP violation while imposing discrete symmetries on a three-Higgs-doublet model \cite{Weinberg:1976hu}. His proposal amounts to imposing a $\mathbb Z_2 \times \mathbb Z_2$ symmetry on the theory allowing for complex coefficients in the potential. The most general renormalizable potential for three doublets $\phi_i$, invariant under $\mathbb Z_2 \times \mathbb Z_2$, is $V = V_2 + V_4$, with\footnote{We follow the notation of Ivanov and Nishi \cite{Ivanov:2014doa}.}
\begin{align}
  V_2&=-[m_{11}(\phi_1^\dagger \phi_1)+m_{22}(\phi_2^\dagger \phi_2)+m_{33}(\phi_3^\dagger \phi_3)]
\end{align}
and
\begin{align}
V_4 &= V_0 + V_\text{ph},\\
V_0 &=\lambda_{11}(\phi_1^\dagger\phi_1)^2+\lambda_{12}(\phi_1^\dagger\phi_1)(\phi_2^\dagger\phi_2)
  +\lambda_{13}(\phi_1^\dagger\phi_1)(\phi_3^\dagger\phi_3)+\lambda_{22}(\phi_2^\dagger\phi_2)^2 \nonumber \\
  &+\lambda_{23}(\phi_2^\dagger\phi_2)(\phi_3^\dagger\phi_3)+\lambda_{33}(\phi_3^\dagger\phi_3)^2 \nonumber \\
  &+\lambda^\prime_{12}(\phi_1^\dagger\phi_2)(\phi_2^\dagger\phi_1)
  +\lambda^\prime_{13}(\phi_1^\dagger\phi_3)(\phi_3^\dagger\phi_1)
  +\lambda^\prime_{23}(\phi_2^\dagger\phi_3)(\phi_3^\dagger\phi_2)\\ 
V_\text{ph} &= \lambda_1(\phi_2^\dagger\phi_3)^2+\lambda_2(\phi_3^\dagger\phi_1)^2 
+\lambda_3(\phi_1^\dagger\phi_2)^2 + \text{ h.c.}
\end{align}
With real coefficients CP is explicitly conserved by the potential. However there is a
region of parameter space \cite{Branco:1980sz} where the potential $V$ admits a CP violating vacuum. In this work we always assume the potential to be real while natural flavour conservation is achieved by letting the doublets and right-handed fermions have the following $\mathbb Z_2 \times \mathbb Z_2$ charges
\begin{subequations} \label{eq:yukawa-structure}
\begin{align}
\phi_1 : (+1,+1) & & \phi_2 : (-1,+1) & & \phi_3 : (+1,-1) \\
u_R : (+1,+1) & & d_R : (-1,+1) & & e_R : (+1,-1).
\end{align}
\end{subequations}
This choice is not unique. The important point is to allow each set of right-handed fermions to couple to only one Higgs doublet.

\section{CP content of the neutral scalars}
In a CP violating model such as the one considered here, the physical neutral scalars are in general not CP eigenstates. Nevertheless, their CP content, i.e., how close they are to being CP odd or CP even, can be probed by studying certain couplings, of which two examples are presented below. In this analysis, the Higgs basis $5\times5$ diagonalization matrix of the neutral scalars mass matrix denoted $O$, plays a central role. 

\subsection{Gauge couplings: $h_ih_jZ$}
\subsubsection{General features}

The trilinear gauge couplings involving the Z boson and two neutral scalars probe the relative CP content of the two scalars. Indeed, in a CP conserving theory, this coupling vanishes unless the two scalars have opposite CP. Letting the $h_ih_jZ$ coupling, in units of $g/(2\cos\theta_W)$ be denoted $P_{ij}$, it is given in terms the Higgs basis diagonalization matrix as \cite{Plantey:2022jdg}
\begin{equation} \label{eq:P_ij}
P_{ij}=(O_{i2}O_{j4}+O_{i3}O_{j5})-(i\leftrightarrow j).
\end{equation}
We will refer to this quantity as the "Z affinity" $P_{ij}$.
A value of unity would correspond to the strength of the $HAZ$ coupling in the CP-conserving 2HDM, whereas a value of zero corresponds to the vanishing $hAZ$ and $hHZ$ couplings in that model.
On the other hand, in the CP non-conserving 2HDM it is given by linear expressions in the rotation matrix \cite{Grzadkowski:2014ada}, here $P_{12}=e_3/v$, $P_{23}=e_1/v$,  and P$_{31}=e_2/v$,  with $e_1^2+e_2^2+e_3^2=v^2$.

For the 3HDM, it satisfies a similar sum rule. By orthogonality, it follows that
\begin{equation}
\sum_j P_{ij}^2=\sum_{j\neq i} O_{j1}^2=1-O_{i1}^2,
\end{equation}
and thus $\sum_{ij} P_{ij}^2=5-1=4$. By antisymmetry, the 5 diagonal elements $P_{ii}$ vanish, and for a random $O$ matrix the non-zero values of $P_{ij}^2$ would have an average of $4/20=0.2$.

The quantity $P_{ij}$ can be either positive or negative. We have performed scans over the potential parameters \cite{Plantey:2022jdg} and studied the "average" Z affinity as the following root-mean-square
\begin{equation}
\Prms{i}{j}=\sqrt{\frac{1}{N}\sum_1^N P_{ij}^2},
\end{equation}
where the sum runs over the $N$ points of the scan.
When the affinity $\Prms{i}{j}$ is large, the $h_ih_j$ pair has a strong coupling to the $Z$, i.e., they have rather different CP content. On the other hand, when the affinity $\Prms{i}{j}$ is low, the two states are close in their CP profiles and have a suppressed coupling to the $Z$. It should be noted however that the affinities $P_{ij}$ are not transitive: If $P_{ab}=P_{ac}$, that is, $h_b$ and $h_c$ are equally ``far'' from $h_a$ in terms of CP content, then it does {\it not} follow that $P_{bc}=0$, i.e., $h_b$ and $h_c$ do not necessarily have the same CP content. The origin of this property lies in the fact that $P_{ij}$ consists of terms each involving both $i$ and $j$.

\subsection{The regime of $U(1)\times U(1)$ symmetry}

When the rephasing sensitive part of the potential $V_\text{ph}$ vanishes, the $\mathbb Z_2 \times \mathbb Z_2$ symmetry is enhanced to a $U(1) \times U(1)$ symmetry. Whenever this symmetry is broken by the Higgs vacuum expectation values (VEVs), there are two massless states in the scalar spectrum, as expected from the Goldstone theorem \cite{Goldstone:1961eq,Goldstone:1962es}. Moreover, when the $U(1)\times U(1)$ symmetry is exact, at the Lagrangian level, the phases of the VEVs can be rotated away and therefore spontaneous CP violation does not occur.
This model has the interesting property that close to this limit, in general it is controlled by only one parameter, $\lambda_1$, since the other couplings in $V_\text{ph}$, $\lambda_2$ and $\lambda_3$, can be related to $\lambda_1$ by using the minimization conditions for the potential.

If the $U(1)\times U(1)$ symmetry is only approximate, i.e., if $\lambda_1$ is small, but finite with respect to all other quartic couplings, then the spectrum contains two light states corresponding to two pseudo-Goldstone bosons. The phenomenology of the model can be conveniently analysed using $\lambda_1$ since it controls both the masses of the $U(1)\times U(1)$ pseudo-Goldstone bosons and the strength of the CP violating effects coming from the Higgs sector. 

\begin{figure}[htb]
\begin{center}
\includegraphics[scale=0.30]{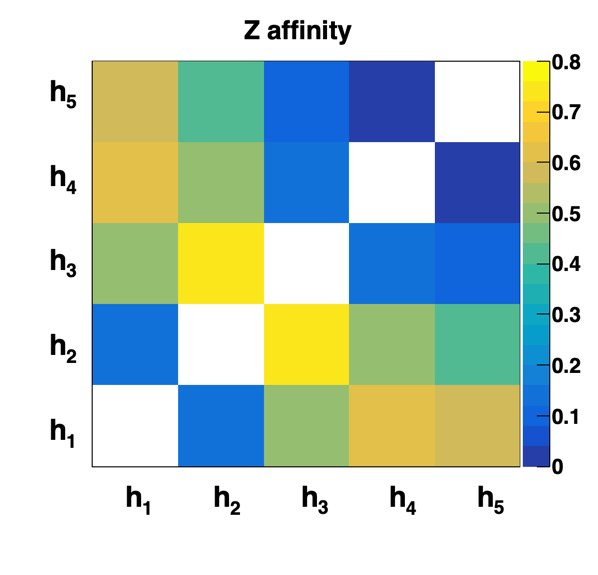}
\includegraphics[scale=0.30]{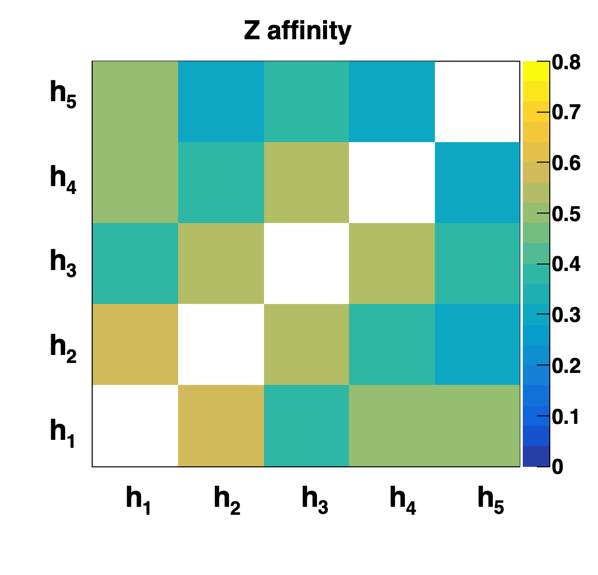}
\end{center}
\vspace*{-4mm}
\caption{Average Z affinity $\Prms{i}{j}$ of states $h_i$ and $h_j$. Left: the $U(1)\times U(1)$ limit, as defined by Eq.~(\ref{Eq:low-lambda}); Right: no restriction on the lambdas.}
\label{Fig:affinity}
\end{figure}

In Fig.~\ref{Fig:affinity} we compare Z affinities for all pairs of neutral scalars, and for two cases. In the left panel, we impose the ``near $U(1)\times U(1)$ symmetry'' condition
\begin{equation} \label{Eq:low-lambda}
\max(|\lambda_1|, |\lambda_2|, |\lambda_3|)=0.01,
\end{equation}
whereas in the right panel we impose no such constraint, i.e., we do not restrict the scan to the regime of near $U(1)\times U(1)$ symmetry. The left panel shows a clear separation into two sets of states, $h_1$ and $h_2$ have low affinity to the $Z$, meaning they have similar CP content, as does the other set, $h_3$, $h_4$ and $h_5$. It is natural to interpret this as follows:
{\it Near the $U(1)\times U(1)$ limit we have two neutral states that are approximately odd under CP, and three that are approximately even.} This is fully compatible with the expectations from the Goldstone theorem \cite{Goldstone:1961eq,Goldstone:1962es}, since the Goldstone bosons in the $U(1)\times U(1)$ limit will be CP odd.

We note that this limit also secures an upper bound on the amount of CP violation in the scalar sector.

\subsubsection{Alignment}
In order to make contact with experimental reality, we have to make sure one scalar has a coupling to the electroweak gauge bosons $W^+W^-$ that is close to unity, in units of $gm_W$. This amounts to requiring $|O_{k1}|$ being close to unity, for some $k$.
Let us first note that in the exact alignment limit, with $|O_{k1}|=1$, then, by unitarity, all $O_{kj}=0$, for $j\neq1$, and all $O_{i1}=0$ for $i\neq k$. Thus,
\begin{equation}
\text{Exact alignment: }\quad P_{ik}=P_{kj}=0, \quad
\text{for all } i \text{ and all }j.
\end{equation}
In this exact alignment limit, the row and column referring to $h_k$ in the $\Prms{i}{j}$ plot would thus be zero, reflecting the fact that $h_j h_k$ would not couple to $Z$ for any $j$.

\begin{figure}[htb]
\begin{center}
\includegraphics[scale=0.30]{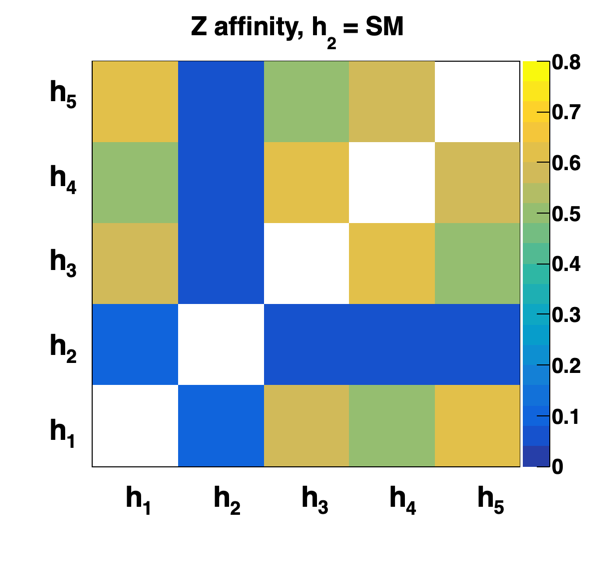}
\includegraphics[scale=0.30]{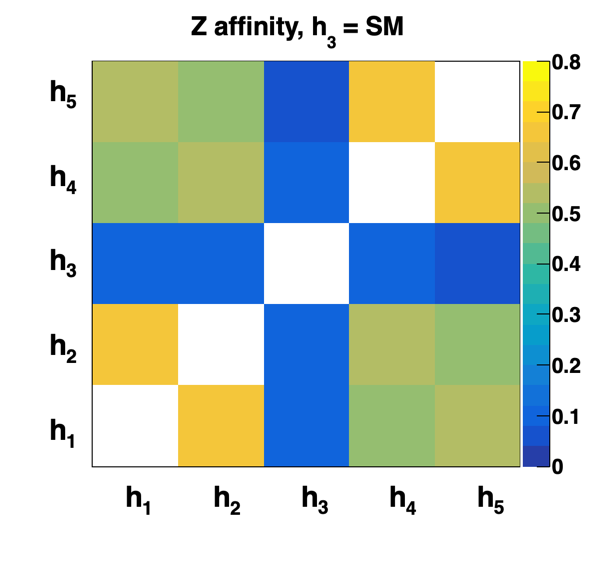}
\end{center}
\vspace*{-4mm}
\caption{Average Z affinity $\Prms{i}{j}$ of states $h_i$ and $h_j$. Left: $h_2$ is assumed to satisfy the SM constraint, Eq.~(\ref{Eq:SM-constraint}); Right: $h_3$ satisfies the SM constraint.}
\label{Fig:affinity-SM}
\end{figure}

We shall relax the alignment condition, replacing it by the experimental one \cite{Zyla:2020zbs}, obeyed at the $3\sigma$ level:
\begin{equation} \label{Eq:SM-constraint}
\kappa_V-3\sigma<O_{k1}^2<\kappa_V+3\sigma,\quad \text{for some }k.
\end{equation}
Imposing this constraint on $h_2$ and $h_3$, the scan \cite{Plantey:2022jdg} yields the average Z affinities shown in Fig.~\ref{Fig:affinity-SM}. Qualitatively, the features described above are verified. Couplings of the $Z$ to the ``SM state'' are strongly suppressed, whereas there can be strong couplings to $h_i h_j$, provided $i$ and $j$ both differ from $k$.

\subsection{Neutral mass spectrum}
In the Weinberg potential with an SM-like Higgs boson state, there is a strong tendency for one or two additional bosons to be lighter than $125$~GeV. In the parameter scan performed in Ref.~\cite{Plantey:2022jdg} the SM state $h_k$ is the second lightest in 38\% of the cases, and the third lightest in an additional 28\% of the cases.
However, this does not necessarily rule out much of the parameter space of the model as the light states may have escaped detection at LEP where the dominant production channel would be the Bjorken mechanism. Indeed, the cross section for this channel is proportional to the $h_iZZ$ couplings $O_{i1}$ which are suppressed for the BSM scalars by the orthogonality of the mixing matrix. 

\subsection{Yukawa couplings}
Even though this model would not generate a complex CKM matrix, 
for real Yukawa couplings there would still be CP violation in the Yukawa sector coming from Higgs interactions. Indeed, the neutral scalars $h_i$ do not have definite CP and have both CP-even and CP-odd couplings to fermions. Hence their Yukawa interactions are given by
\begin{align}
\mathcal L_{h_iff} = \frac{m_f}{v}h_i(\kappa^{h_iff}\bar f f + i\tilde\kappa^{h_iff} \bar f \gamma_5 f).
\end{align}
One can define the angle $\alpha^{h_iff}$ through 
\begin{equation}
\tan \alpha^{h_iff} = \frac{\tilde \kappa^{h_iff}}{\kappa^{h_iff}}.
\end{equation}
which measures the relative strength of the CP-even and CP-odd coupling of $h_i$ to a fermion $f$ and hence also measures the CP profile of $h_i$.
In this model, when $f$ couples to the doublet $\phi_k$, $\alpha^{h_iff}$ can be computed in terms of the matrix $\tilde{\cal R}$  \cite{Plantey:2022jdg} which rotates the doublets to the Higgs basis and the rotation matrix $O$ which diagonalizes the neutral mass-squared matrix in the Higgs basis
\begin{align}
\alpha^{h_iff} = \arg(Z_i^{(k)})
\end{align}
where
\begin{align}
Z_i^{(k)}&=\left(\tilde{\cal R}^T\right)_{k1}O_{i1}+\left(\tilde{\cal R}^T\right)_{k2}(O_{i2}+iO_{i4})
+\left(\tilde{\cal R}^T\right)_{k3}(O_{i3}+iO_{i5}) \nonumber \\
&=\tilde{\cal R}_{1k}O_{i1}+\tilde{\cal R}_{2k}(O_{i2}+iO_{i4})
+\tilde{\cal R}_{3k}(O_{i3}+iO_{i5}).
\end{align}
However, it  should be pointed out that a real CKM matrix is already ruled out by experiment \cite{Botella:2005fc,Charles:2004jd}. The simplest way to fix this problem is to allow for complex Yukawa couplings.

\section{Conclusion}
In most of the parameter space of the Weinberg potential, the scalar spectrum contains at least one neutral scalar lighter than the observed Higgs boson. This is not necessarily a problem for the model because the main production channel for these light states would have been suppressed due to the orthogonality of the scalar mass mixing matrix. In fact there has been recent interest in light Higgs bosons due to observed excesses in Higgs searches around $95$ GeV \cite{LEPWorkingGroupforHiggsbosonsearches:2003ing, CMS:2018cyk, CMS:2015ocq, Biekotter:2022jyr}. It would be interesting to explore if a model based on Weinberg's 3HDM potential could fit these excesses, while obeying current experimental constraints.

\bibliography{ref}{}
\bibliographystyle{JHEP}

\end{document}